\documentclass[prl,onecolumn,amsmath,amssymb,superscriptaddress,showpacs]{revtex4}

\usepackage{graphicx}
\usepackage{epsfig}
\usepackage{epstopdf}

\usepackage{import}
\usepackage{color}

\begin{document}

\title{Biexciton, single carrier, and trion generation dynamics in single-walled carbon nanotubes}
\author{B.~Yuma}
\affiliation{IPCMS, CNRS and Universit\'e de Strasbourg, 23, rue du L{\oe}ss, F-67034 Strasbourg, France} 
\author{S.~Berciaud}
\affiliation{IPCMS, CNRS and Universit\'e de Strasbourg, 23, rue du L{\oe}ss, F-67034 Strasbourg, France} 
\author{J.~Besbas}
\affiliation{IPCMS, CNRS and Universit\'e de Strasbourg, 23, rue du L{\oe}ss, F-67034 Strasbourg, France} 
\author{J.~Shaver}
\affiliation{LP2N, Universit\'e de Bordeaux, Institut d'Optique Graduate School, and CNRS, 351 cours de la Lib\'eration, F-33405 Talence, France} 
\author{S.~Santos}
\affiliation{LP2N, Universit\'e de Bordeaux, Institut d'Optique Graduate School, and CNRS, 351 cours de la Lib\'eration, F-33405 Talence, France} 
\author{S.~Ghosh}
\affiliation{Department of Chemistry and R.E. Smalley Institute for Nanoscale Science and Technology, Rice University, 6100 Main Street, Houston, Texas 77005, USA} \author{R.B.~Weisman}
\affiliation{Department of Chemistry and R.E. Smalley Institute for Nanoscale Science and Technology, Rice University, 6100 Main Street, Houston, Texas 77005, USA}
\author{L.~Cognet}
\affiliation{LP2N, Universit\'e de Bordeaux, Institut d'Optique Graduate School, and CNRS, 351 cours de la Lib\'eration, F-33405 Talence, France}
\author{M.~Gallart}
\affiliation{IPCMS, CNRS and Universit\'e de Strasbourg, 23, rue du L{\oe}ss, F-67034 Strasbourg, France} 
\author{M.~Ziegler}
\affiliation{IPCMS, CNRS and Universit\'e de Strasbourg, 23, rue du L{\oe}ss, F-67034 Strasbourg, France} 
\author{B.~H{\"o}nerlage}
\affiliation{IPCMS, CNRS and Universit\'e de Strasbourg, 23, rue du L{\oe}ss, F-67034 Strasbourg, France} 
\author{B.~Lounis}
\affiliation{LP2N, Universit\'e de Bordeaux, Institut d'Optique Graduate School, and CNRS, 351 cours de la Lib\'eration, F-33405 Talence, France} 
\author{P.~Gilliot}
\affiliation{IPCMS, CNRS and Universit\'e de Strasbourg, 23, rue du L{\oe}ss, F-67034 Strasbourg, France} 
\email{pierre.gilliot@ipcms.unistra.fr}   

\date{\today}

\begin{abstract}
We present a study of free carrier photo-generation and multi-carrier bound states, such as biexcitons and trions (ionized excitons), in semiconducting single-walled carbon nanotubes. Pump-and-probe measurements performed with fs pulses reveal the effects of strong Coulomb interactions between carriers on their dynamics. Biexciton formation by optical transition from exciton population results in an induced absorption line (binding energy 130~meV). Exciton-exciton annihilation process is shown to evolve at high densities towards an Auger process that can expel carriers from nanotubes. The remaining carriers give rise to an induced absorption due to trion formation (binding energy 190~meV). These features show the dynamics of exciton and free carriers populations.
\end{abstract}


\maketitle

\section{{Introduction}}

The excitonic nature of optical resonances in single-walled carbon nanotubes (SWCNTs) \cite{wang_optical_2005,maultzsch_exciton_2005} has stimulated numerous research studies on the influence of Coulomb interaction on the photophysics  of these quasi one-dimensional systems. Enhanced Coulomb interaction and reduced dielectric screening in SWCNTs give rise to considerable many-body effects/bandgap renormalization and binding energies of excitons ($X$), as high as $700~\rm meV.nm$ for freestanding semiconducting SWCNTs\cite{capaz_diameter_2006, lefebvre_excited_2008}, a value that is a large fraction of the single particle bandgap.
Linear optical spectroscopy has revealed the richness of the exciton manifold~\cite{bachilo_2002, dresselhaus_exciton_2007}. For sufficiently high excitation rates, many body bound states such as  biexcitons\cite{kammerlander_biexciton_2007, colombier_2012, pedersen_2005, watanabe_2011} ($XX$) or charged excitons \cite{ronnow_correlation_2010,watanabe_trion_2012} (trions, $X^*$) should form. Because of the strong Coulomb interaction between photogenerated charge carriers, such SWNT states have been predicted to have significant binding energies in the 60 - 250~meV range for biexcitons~\cite{kammerlander_biexciton_2007}.

Biexcitons are bound states of two excitons similar to a hydrogen molecule. They usually arise from the collision between two excitons, but they can be also generated from an exciton population by one-photon absorption, resulting in an induced absorption line~\cite{Levy_1985,BawendiPRL1990,HuPRL1990}. In carbon nanotubes, however, Coulomb interaction also results in highly efficient Auger processes such as exciton-exciton annihilation  (EEA)~\cite{wang_observation_2004,huang_quantized_2006,valkunas_exciton-exciton_2006}, that are expected to bypass biexciton formation. This is in agreement with the complete saturation behavior reported in air-suspended nanotubes~\cite{xiao_saturation_2010}. 

At high exciton densities, Coulomb interactions may also lead to exciton ionization and the subsequent generation of a carrier gas, whose signature is the appearance of a new transition toward trion states. 
Trions have been first predicted and observed in inorganic semiconductors~\cite{Kheng1993}. They can be seen as the bound state of three carriers: an optically excited electron-hole (e-h) pair and an additional carrier (electron or hole) that is already present. This simple picture works mainly for strongly confined zero-dimensional systems (\textit{e.g.} quantum dots), that can only accommodate a few carriers. In extended systems (\textit{e.g.} 2-dimensional quantum wells), the trion is one of the optical excitations of a many-body system, corresponding to the creation of an e-h pair within a dense carrier gas that is usually obtained by doping~\cite{Kheng1993, Finkelstein_1995, finkelstein_shakeup_1996}. Nevertheless, in quantum wells, trions have been shown~\cite{brinkmann1999, gilliot1999} to be mainly created on localized carriers that are trapped on electrostatic potential fluctuations. This more realistic situation for extended systems, with an e-h pair bound to a single localized carrier, is thus very close to the case of quantum dots. In any event, trion formation requires that charge carriers are already present in the excited structure, either by doping or through the separation of the electron and hole forming optically excited pairs. Trion optical absorption and emission act then as a probe of the population and localization of the carrier gas. In SWCNTs, the impact of the local electrostatic environment on the photophysical properties has recently been put forward by our demonstration of all-optical trion generation in pristine SWCNTs~\cite{Santos2011}. For sufficiently high laser intensities, a red-shifted trion line could be observed in both photoluminescence (PL) and transient absorption experiments. Similar trion features were also reported in a static PL study by Matsunaga \textit{et al.} on intentionally hole-doped nanotube ensembles~\cite{matsunaga_observation_2011}. In that case, the presence of the additional carriers allowing trion formation was expected, whereas the observation of charged many-body bound states in pristine samples through carrier photogeneration was singular and less anticipated~\cite{Santos2011}. Interestingly, no evidence of bi-exciton formation could be found in the emission spectra.

In this letter, we report a detailed transient absorption study that allows to characterize the dynamics of the processes that result from Coulomb interactions: many-body bound state formation and carrier generation originating from exciton collisions. Our data show evidence for biexciton \textit{and} free carrier formation in chirality sorted (6,5) SWCNTs. Following the generation of several excitons by a pump pulse, a probe pulse allows the observation of two induced absorption (IA) features. These are red-shifted with respect to the main induced transmission (IT) feature assigned to the $\rm S_{11}$ exciton and are attributed to the formation of biexcitons and trions. We show that the $X$ IT and $X^*$ IA dynamics reflect the evolutions of the exciton and free carrier populations, respectively. From analysis of these dynamics, we deduce the processes that allow photogeneration of carriers through exciton-exciton collisions and their spatial separation.

\section{{Experimental setup and samples}}
\begin{figure}[!htb]
\begin{center}
\includegraphics[width=9cm]{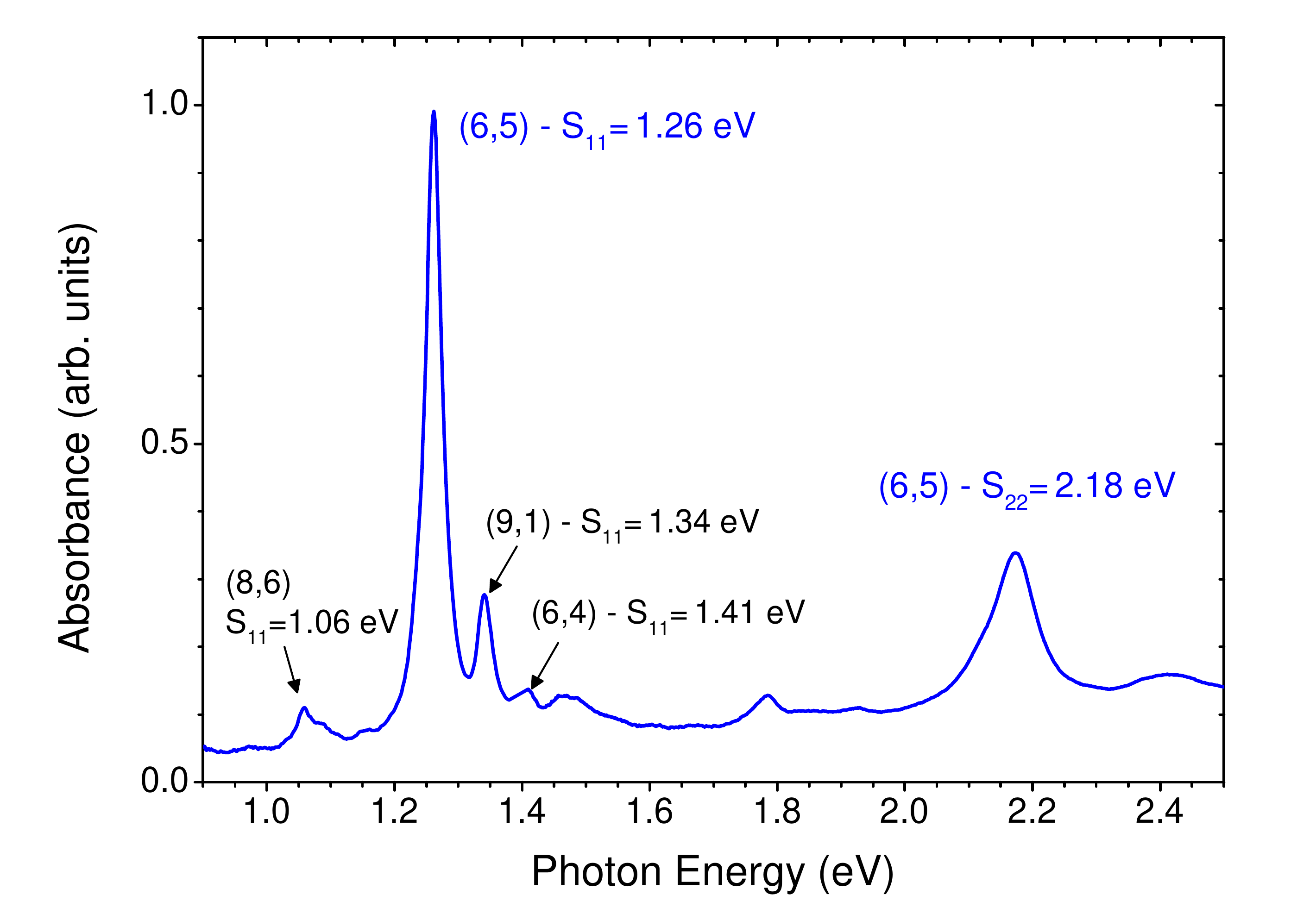}
\caption{Linear absorption spectrum of enriched (6,5) SWCNTs. The $\rm S_{11}$ exciton and $\rm S_{22}$ exciton of (6,5) SWCNTs, as well as contributions from residual species, are indicated by arrows.}
\label{AbsLin}
\end{center}
\end{figure}

Since the characteristic timescales associated with EEA processes are in the picosecond range~\cite{wang_observation_2004,huang_quantized_2006,valkunas_exciton-exciton_2006}, femtosecond pump-probe spectroscopy is the method of choice to study optical excitations in nanotubes. Such pump-probe experiments have often been performed on heterogeneous ensembles of SWCNTs composed of many different species. In these conditions, experimental fingerprints from many-body bound states may be masked by unwanted residual signals from non-resonantly excited species. To overcome this issue, we performed transient absorption spectroscopy measurements on an ensemble of chirality-sorted (6,5) SWCNTs. Enrichment in the (6,5) chirality was obtained by nonlinear density gradient ultracentrifugation (DGU)~\cite{ghosh_advanced_2010}. The nanotubes are dispersed in an aqueous solution of sodium cholate in a fused silica cell. As shown in Fig.~\ref{AbsLin}, the linear absorption spectrum is largely dominated by contributions from (6,5) SWCNTs.

For the time-resolved measurements, we used an amplified Ti:Sapphire laser that produces ultrashort pulses (100 fs) at a repetition rate of $200~\rm kHz$. The energy of pump photons was tuned into resonance with the $\rm S_{11}$ or $\rm S_{22}$ transitions (X lines) at $1.26~eV$ or $2.19~eV$, respectively, by means of an optical parametric amplifier (OPA). The pump fluence range corresponds to $5\times 10^{12} - 2.5\times 10^{14}$ photons per $\rm cm^2$. Ultrashort continuum probe pulses were generated in a sapphire crystal. Their spectral range (0.95-1.35~eV) covered the region where the exciton, trion, and biexciton lines are observed.  Transmission spectra of the probe beam were acquired as a function of the time-delay between the pump and probe pulses using a InGaAs linear diode array at the output of a spectrometer, and differential transmission spectra were calculated from the data. In order to study the dynamics of the various spectral features, we extracted their contributions as a function of pump-probe delay.

\section{{Transient absorption spectra}}

\begin{figure}
\includegraphics[width=8.5cm]{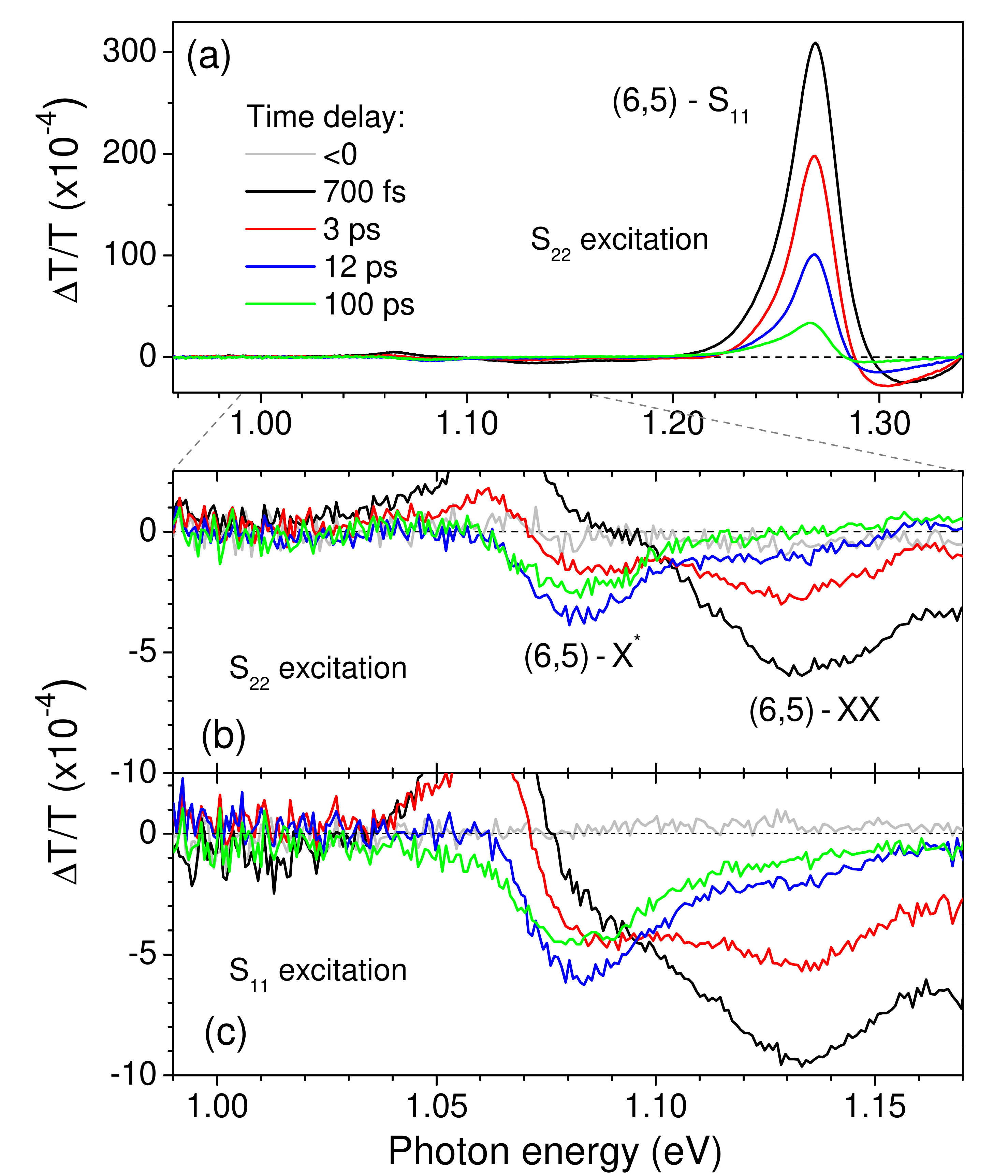}
\caption{(a)-(b) Differential transmission spectra obtained after excitation at $\rm S_{22}$ excitonic transition. The IT at 1.26~eV is the signature of the $\rm S_{11}$ excitonic transition from (6,5) nanotubes. The pump fluence is $\rm 6.3\times 10^{13} {\rm photons}/{\rm pulse}/{\rm cm}^2$.(b) Zoom on the spectral region below the energy of the exciton of the (6,5) nanotube. At 1.06~eV, an IT is present at very short times and attributed to the $\rm S_{11}$ excitonic transition from a residual fraction of (8,6) nanotubes. The IA at 1.08~eV is attributed to trion ($X^{*}$) formation from a charge level. At 1.13~eV, another IA, attributed to biexciton ($XX$) formation is visible from short delays up to 50~ps approximately. (c) Differential transmission spectra in the same near infrared region after excitation at the $\rm S_{11}$ excitonic transition. The pump fluence is $\rm 1.2 \times 10^{14} {\rm photons}/{\rm pulse}/{\rm cm}^2$.} 
\label{FigPeaks}
\end{figure}

Figure~\ref{FigPeaks} shows differential transmission spectra that were obtained after excitation by pump pulses tuned to 2.19~eV near the $\rm S_{22}$ excitonic absorption line of (6,5) nanotubes.  From comparison with the linear absorption spectrum, the strong IT peak that appears in Fig.~\ref{FigPeaks}(a) at 1.26~eV is attributed to the $\rm S_{11}$ excitonic transition from (6,5) nanotubes. Three other features are also observed in the near-infrared region that lies below, as shown in Figure~\ref{FigPeaks}(b): an IT at $\sim$~1.06~eV and two IA at 1.08~eV and 1.13~eV. 

Similar pump and probe measurements were performed using resonant pumping of the (6,5)-$\rm S_{11}$ exciton line at 1.26~eV, with a laser fluence of $1.2 \times  10^{14} \rm {photons/pulse/cm^2}$. Fig.~\ref{FigPeaks}(c) presents the corresponding differential transmission spectra. We observed no major spectral differences from the spectra obtained by pumping at $\rm S_{22}$. These similar behaviors are due to the extremely fast ($\sim 40 {\rm fs}$) $\rm S_{22}$ to $\rm S_{11}$ intersubband decay\cite{manzoni_intersubband_2005} which is not resolved in our measurements. 

\section{{Exciton bleaching and dynamics}}
The (6,5)-exciton bleaching at 1.26~eV can be attributed to absorption saturation by the pump beam that gives rise to an IT feature in the probe spectrum. It is due to both phase-space filling (PSF) and Coulomb interaction (CI) between excitons~\cite{Shah1999,Haug1984,luer_size_2009,NguyenPRL2011}. On the high energy side of the $\rm S_{11}$ exciton line, at 1.3~eV, an IA rises together with the IT. We suggest that this blue-shifted IA arises from the strong exciton-exciton interaction that occurs in one-dimensional systems. In particular, screening of the electron-hole interaction within the high density exciton gas leads to a decrease of their binding energy and therefore to a blue shift of the X line. As seen in figure~\ref{FigPeaks}, the exciton IT signal rises rapidly, within a few hundred femtoseconds. This time scale is close to the temporal resolution of our setup. The signal mostly decreases within $\sim 10~\rm ps$, although there remains a slowly decaying contribution after $100 ~\rm ps$~\cite{zhu_pump-probe_2007,Nishihara_2012}. This fast initial decay has been shown to result from exciton-exciton annihilation (EEA)~\cite{wang_observation_2004,huang_quantized_2006,valkunas_exciton-exciton_2006} that destroys the exciton population within a few picoseconds.
The dynamics of the IT observed at $\sim$~1.06~eV is very similar to that observed for the (6,5) exciton and described above. According to the linear absorption spectrum (cf. Fig.\ref{AbsLin}), this IT is attributed to the (8,6)-$\rm S_{11}$ exciton transition. It will not be discussed further but has been taken into account when fitting the differential transmission spectra.

\section{{Biexciton formation}}
We now consider the IA feature observed at 1.13~eV, which is red-shifted by about 130~meV from the $S_{11}$ bright exciton IT feature. This spectral position coincides with the emission of the (6,5)-$\rm S_{11}$ phonon sideband that couples a K-momentum dark exciton to near zone edge optical phonons~\cite{matsunaga_origin_2010, vora_chirality_2010}. However, the signal observed in our pump-probe experiments is an induced absorption, whereas can emission would appear as an induced transmission. Thus, it cannot be the fingerprint of a phonon sideband. Another possibility would be the emission of a brightened triplet exciton state appearing in photo-damaged SWCNTs, as observed under strong laser irradiation~\cite{matsunaga_origin_2010}. Such defect-induced processes should however appear in the linear absorption spectra, which is not the case in our experiments. In addition, their signature in differential transmission measurements should also be an IT.

\begin{figure}[!htb]
\begin{center}
\includegraphics[width=7.5cm]{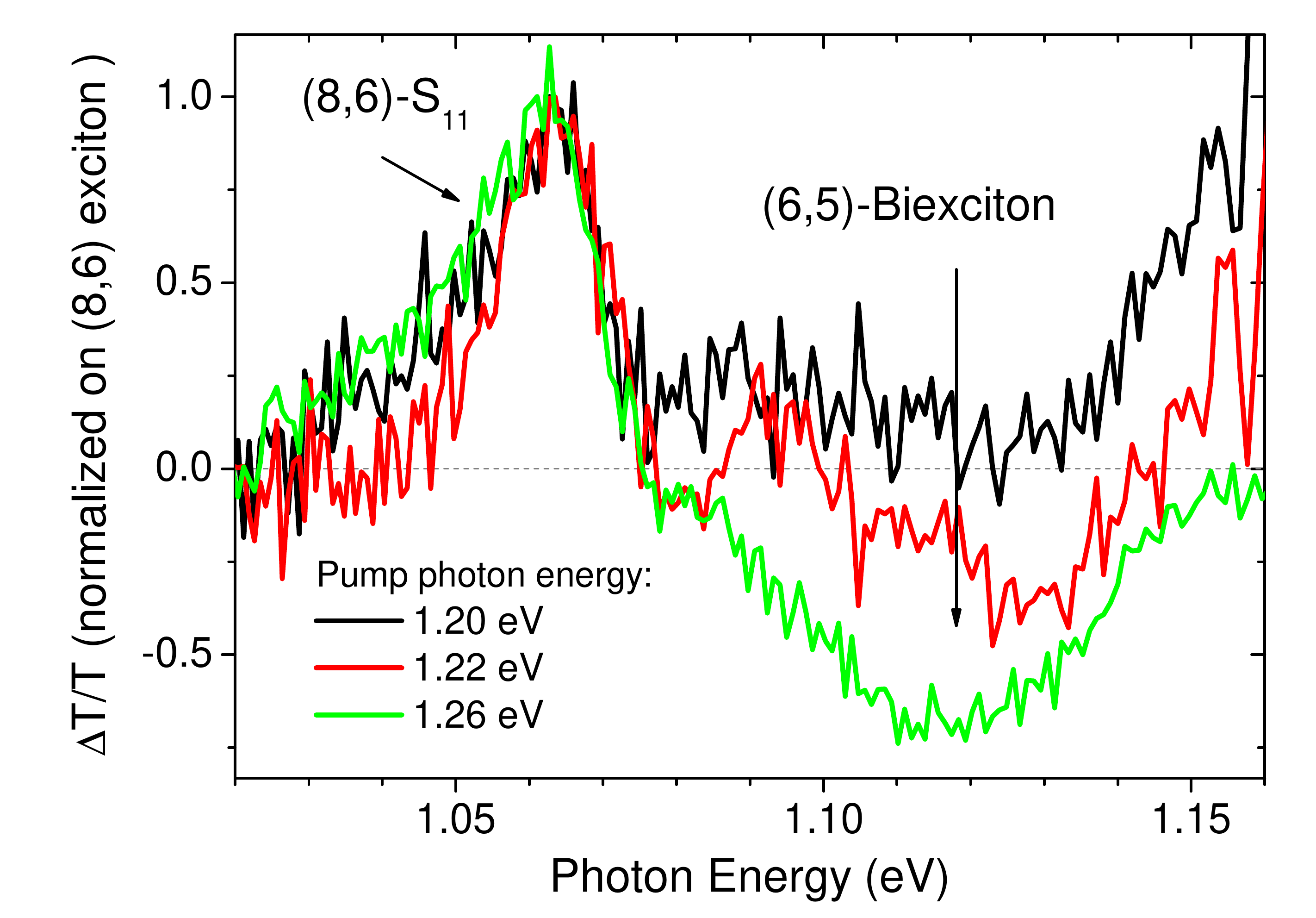}
\caption{
Transient spectra showing IT at 1.06 eV ((8,6)-SWCNT exciton) and IA at $1.13\rm eV$ ((6,5)-exciton-biexciton transition) after excitation at different photon energies, close to the $\rm S_{11}$ line. The delay between pump and probe pulses is fixed at $700 \rm fs$. The spectra have been normalized to the exciton signal for ease of comparison.
}
\label{mecaBiexciton}
\end{center}
\end{figure}

\begin{figure}[!htb]
\begin{center}
\includegraphics[width=8.5cm]{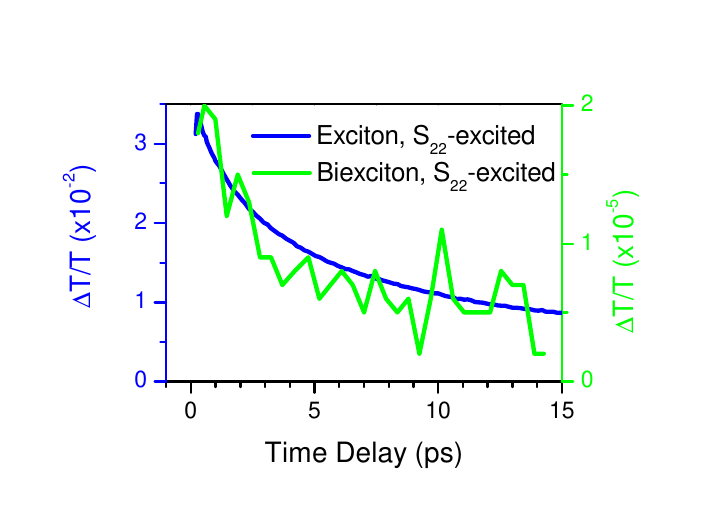}
\caption{
Normalized transient spectra of $\rm S_{11}$ exciton IT at $1.26~eV$ (blue curve), and biexciton IA at 1.13~eV (green line). The IA have been plotted using a reversed Y-axis for ease of comparison with the IT of the exciton. The laser fluence was $2.7 \times 10^{13} {\rm photons}/{\rm pulse}/{\rm cm}^2$. The excitation was performed on the $\rm S_{22}$ transition.
}
\label{FigXX2}
\end{center}
\end{figure}

A more compelling interpretation for this new IA feature is the creation by the probe pulse of a biexciton $(XX)$, from the exciton population previously generated by the pump pulse, \textit{i.e.} to the $\rm X\rightarrow \rm XX$ optical transition. The separation of 130~meV observed in our experiments between the line at 1.13~eV and the exciton line at $1.26~eV$ is large compared to thermal energy. Its magnitude order matches theoretical calculations~\cite{kammerlander_biexciton_2007}. To verify that this feature arises from (6,5) SWCNTs, we performed differential transmission measurements using different pump photon energies at and below the $\rm S_{11}$ transition. The spectra shown in Fig.~\ref{mecaBiexciton} are normalized with respect to the amplitude of the IT feature due to the residual population of (8,6) SWCNTs at a given time delay. While the exciton IT signal of the (8,6) nanotubes is not expected to change, as they are excited above the transition, the amplitude of the $XX$ IA feature decreases strongly when the energy of the pump photons is tuned below the $S_{11}$ resonance of (6,5) SWCNTs and the photoinduced population of excitons in those nanotubes becomes smaller. 

The transient dynamics of the $XX$ feature, represented in Figure~\ref{FigXX2} along with that of the $X$ IT, bolsters our interpretation.
The dynamics of the $X$ and $XX$ features are quite similar, with a signal decay that occurs mainly in the first 5~ps. Then, even if the exciton population undergoes EEA, the corresponding IA absorption should be observed, as long as an exciton population remains. The signal dynamics should thus follow that of the exciton population, and not the biexciton state lifetime, which may be very short. Measurements performed at higher pump fluence (not shown here) show a faster decay of the $XX$ IA feature that follows the enhanced exciton decay due to EEA.
We emphasize that our observation is not contradictory with previous PL experiments~\cite{matsuda_exciton_2008, Santos2011} that have ruled out the formation of biexcitons because of the competition with highly efficient EEA. Indeed, the process that involves a collision between two excitons to form a biexciton, is highly improbable if those collisions also give exciton-exciton annihilation with a much higher probability.

\section{{Single charge carrier generation and trion formation}}
\begin{figure}[!htb]
\begin{center}
\includegraphics[width=8.5cm]{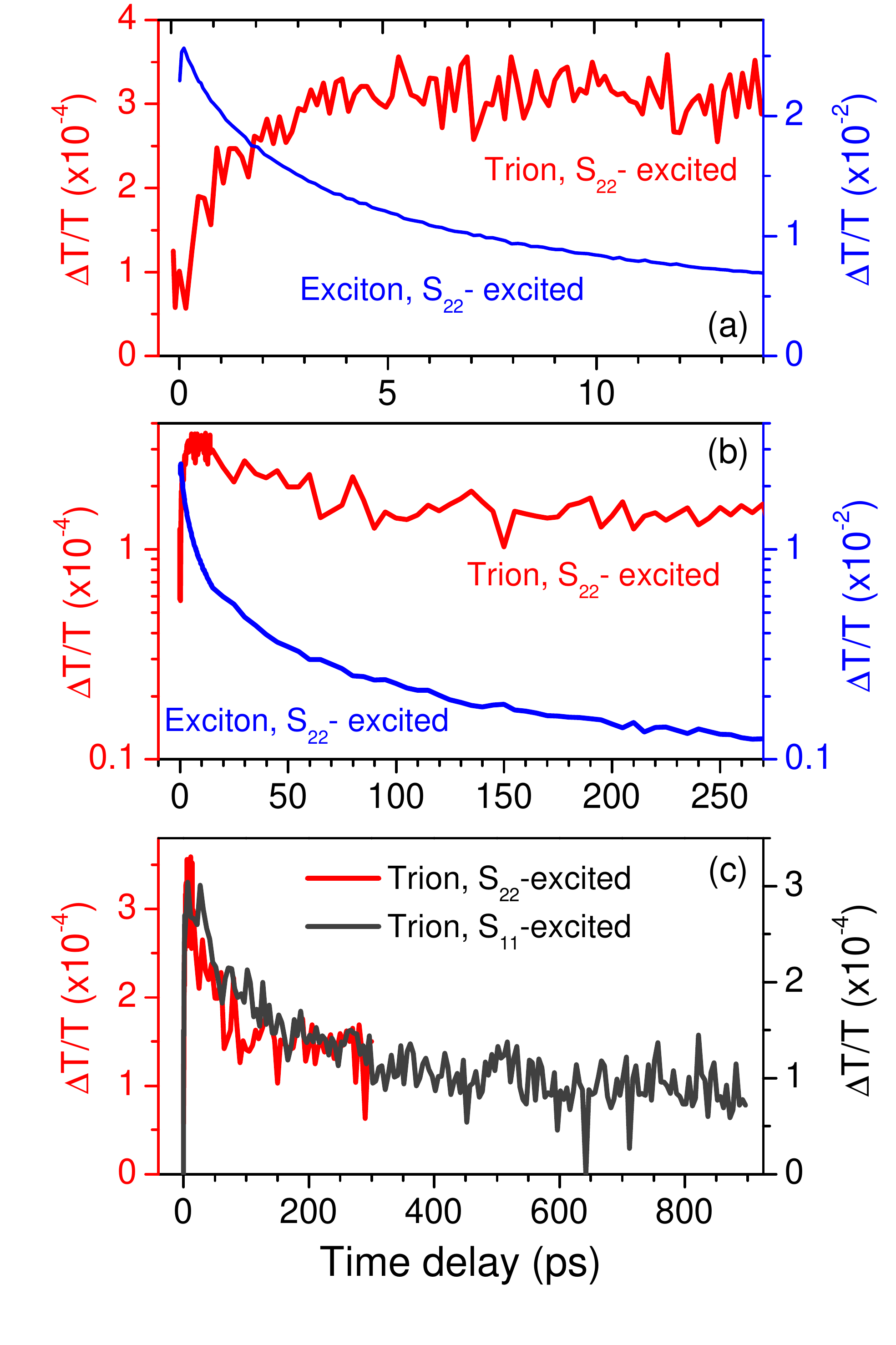}
\caption{
(a) and (b) Normalized transient $\rm S_{11}$ exciton IT at $1.26~eV$ (blue line) and trion IA at $1.08~eV$ (red line). The IA have been plotted using a reversed Y-axis.   The excitation is performed on the $S_{22}$ transition. The laser fluence is $1.3 \times 10^{13} {\rm photons}/{\rm pulse}/{\rm cm}^2$. 
(c) Normalized transient IA signal at the trion transition after excitation at 1.26~eV ($\rm S_{11}$) (gray line) and after excitation at 2.19~eV ($\rm S_{22}$) (red line). The IA have been plotted using a reversed Y-axis. The laser fluence for the $S_{11}$ excitation is $1.16 \times 10^{13} {\rm photons}/{\rm pulse}/{\rm cm}^2$.
} 
\label{FigTrion}
\end{center}
\end{figure}

As explained in our previous report~\cite{Santos2011}, comparison with PL spectra of an ensemble of intentionally doped SWCNTs~\cite{matsunaga_observation_2011} or pristine single SWCNT under intense laser excitation demonstrates that the IA at 1.08~eV can be attributed to trion ($\rm X^*$) formation from a charge carrier level, with a binding energy of 190~meV. Remarkably, as shown in Fig.~\ref{FigTrion}(a), the trion IA rises within $\sim$5~ps. This timescale is very close to the characteristic decay time of the $S_{11}$ excitonic IT feature. This shows that there is a common mechanism responsible for the decay of the exciton and the rise of the trion feature. This supports the following scheme~\cite{Santos2011}: exciton-exciton collisions, which are responsible of the decay of their population at short delays, lead to the formation of a population of charge carriers that is evidenced by the induced absorption toward the trion state. We note that the sign of the trion cannot be determined here: since electron and holes in nanotubes have very similar effective masses, the binding energies of positive and negative trions are expected to be nearly equal, giving the same spectral signature for negative and positive trions~\cite{Park_2012}.

\begin{figure}[!htb]
\begin{center}
\includegraphics[width=8.5cm]{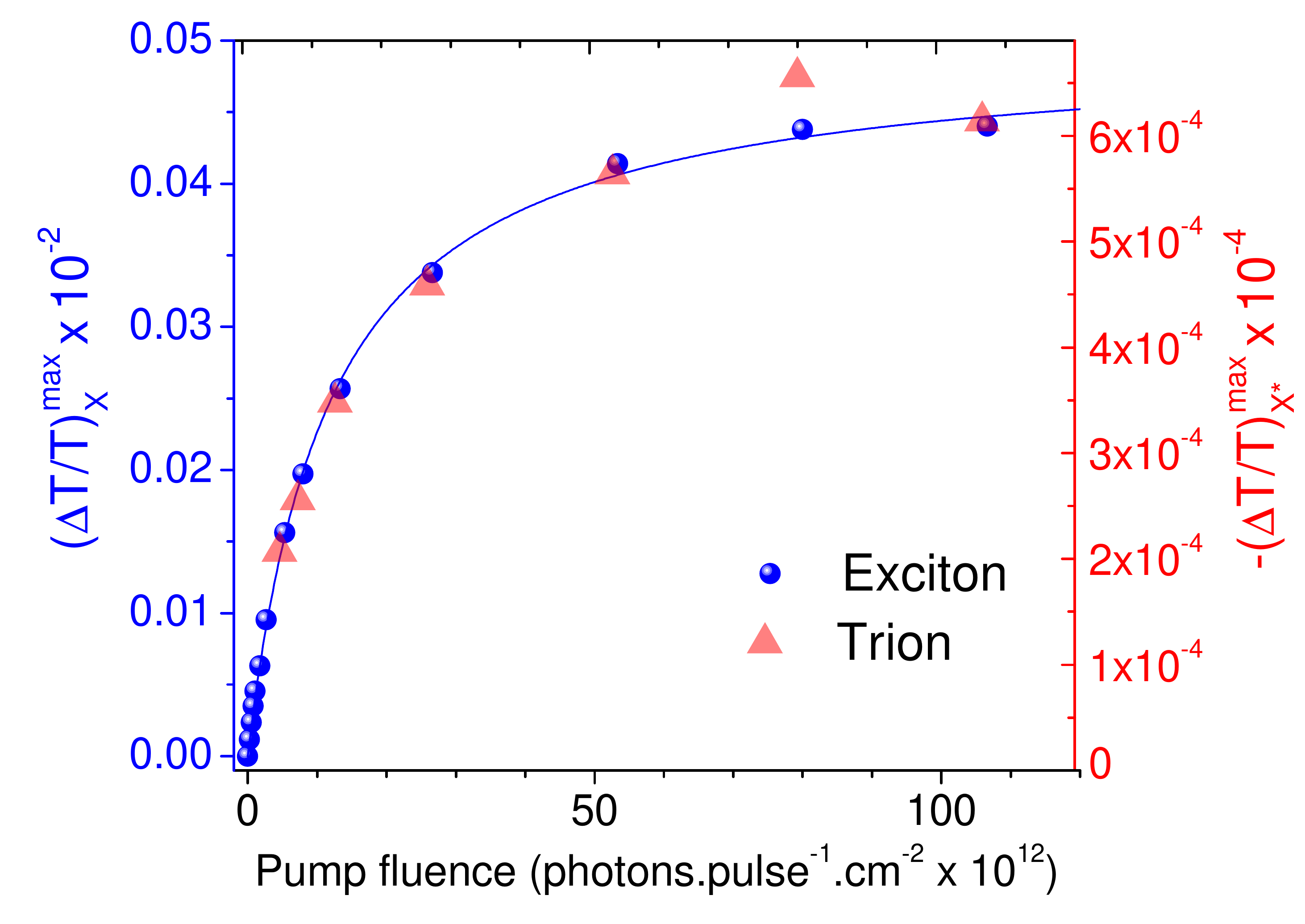}
\caption{Maximum of the IA signal at the photon energy of the trion (red triangles, right Y-axis) and maximum of the IT signal at the photon energy of the exciton (blue circles, left Y-axis) versus pump fluence.}
\label{FigXTnorm}
\end{center}
\end{figure}

We now turn to a more quantitative analysis of our data and first examine the impact of the interplay between the exciton and charge carrier populations on the differential transmission signals. As mentioned above, in the absence of many-body bound states, the existence of the $S_{11}$ exciton IT feature can be understood in terms of phase space filling (PSF) and Coulomb interactions (CI) between excitons~\cite{luer_size_2009}. Here, we assume that some of the photoexcited e-h pairs dissociate to produce single carriers that assist trion formation. Consequently, part of the exciton oscillator strength is transferred to the trion transition. This effect should result in an IT (absorption decrease) at the $X$ photon energy together with an IA at the $X^*$ photon energy. On the other hand, because trions and excitons share the same electron and hole band states from which their wave function is built, they should undergo the same saturation effects (PSF and CI), and the $X^*$ IA amplitude should diminish, while the $X$ IT amplitude should increase. Strictly speaking, the saturation effects and the oscillator strength transfer are thus involved at once in both $X$ and $X^*$ differential transmission. But their respective impact can be evaluated through the maximum amplitude of the signals. As shown in Fig.~\ref{FigPeaks} and~\ref{FigXTnorm}, the differential transmission of the exciton does not exceed 5\%, while the trion IA is two orders of magnitude smaller, limited to 0.07\%.
Thus, the $X^*$ signal is mainly due to oscillator strength transfer and the $X$ signal originates essentially from e-h population effects.
Consequently, one can reasonably consider that the $X$ signal amplitude measures the exciton density $n_X$, while the $X^*$ signal is proportional to the density $n_c$ of single charge carriers.
\begin{eqnarray}
{\left.{\Delta T\over T}\right|}_X \propto n_X (t) & & {\left.{\Delta T\over T}\right|}_{X*} \propto n_c (t)
\label{eq1}
\end{eqnarray}
Nevertheless, the $X$ IT signal shows a component that persists after 200~ps, with an amplitude roughly three times larger than that of the $X^*$ IA ( (see Fig.~\ref{FigTrion}(b)). As evidenced in static absorption of doped nanotubes~\cite{Nishihara_2012}, a population of carriers induces an exciton bleaching that is not completely balanced by the transfer of oscillator strength towards the trion transition. This shows that part of the exciton IT signal, observed at times significantly longer than the exciton lifetime (that is in the range~\cite{hagen_exponential_2005,berciaud_luminescence_2008,gokus_mono-_2010} of 10 ps to 100 ps), is due to band filling effects by carrier populations that decay slowly, as shown below.

Figure~\ref{FigXTnorm} represents the maximum induced transmission of the $X$ feature, which is proportional to the exciton number $n_X^0$, initially created by the pump pulse, along with the maximum induced absorption of the $X^* $ feature, that is attained after a few ps (cf. Fig. \ref{FigTrion}a) and is proportional to the maximum density of charge carriers $n_c^{0}$. They scale identically as a function of the pump fluence (\textit{i.e.} as a function of the number of excitons). This could seem surprising, considering that the derivative of the population $n_c$ of single carriers results from Auger processes and thus scales quadratically  with the exciton population $n_X$. However, since the creation of carriers results from the annihilation of excitons, the same term appears, with an opposite sign, in the differential rate equations that describe exciton decay and the photo-induced generation of charge carriers, respectively. At times shorter than the exciton lifetime, where recombination processes other than EEA can be neglected, this results in:
\begin{equation}
{d n_{c}\over dt} \propto -{d n_{X}\over dt}
\label{constante}
\end{equation}
which gives after integration $n_c^{0} \propto  n_X^{0}$. 
This implies that regardless of the number of prepared excitons $n_X^0$ (or equivalently the laser fluence), one should find a density $n_c^{0}$ that is proportional to $n_X^0$ (as observed in Fig.~\ref{FigXTnorm}). These conclusions hold not only for linear or quadratic processes, but also for any other nonlinear transfer law that connects the two populations $n_c$ and $n_X$ and in particular if there is a threshold for the carrier generation from the excitonic population.  

We can determine the actual initial exciton number per unit length $n_X^0\propto \left.\Delta T/T\right|_{max}$ as a function of the pump fluence using the absorption cross-section value of $1\times10^{-17}{\rm cm}^2$ per carbon atom~\cite{berciaud_luminescence_2008} at low excitation. As shown in Fig.~\ref{FigXTnorm}, it shows a saturation behavior, with a maximum exciton density of $55 \rm \mu m^{-1}$, \textit{i.e.} a separation of $18 ~\rm nm$ between two excitons. This distance can be compared to the exciton Bohr radius~\cite{wang_optical_2005,maultzsch_exciton_2005} in (6,5) SWCNT $a_B=1.2 \rm nm$. Even if the Mott density is not reached, the maximum exciton density lies within one order of magnitude of this limit. At these high population levels, the exciton collision should be strongly modified. In fact, the threshold value $5 \rm \mu m^{-1}$ at which EEA begins to saturate matches the threshold at which the trion signal is observed (Fig.~\ref{FigXTnorm}). It is even more remarkable that the threshold for observing trion emission in PL experiments performed with a cw excitation on single nanotubes is of the same order~\cite{Santos2011}. In these experiments, the trion line emerges from the background at $N_X\simeq 100\rm\mu m^{-1}.ns^{-1}$ that gives, for an exciton lifetime of 100~ps, a mean density $n_X\simeq 10\rm\mu m^{-1}$. As mentioned above, trion formation implies the creation of single carriers following exciton ionization. We thus propose that for densities larger than a few tenths of an exciton per micron, where trions start to be observed in transient absorption experiments, the mechanism involved during an exciton-exciton collision changes and gives other educts. While EEA results in the transfer of the remaining exciton as a whole, for higher $X$ densities, when Coulomb interaction within the exciton is strongly screened, exciton-exciton collision would ionize the remaining exciton, transferring the electron or the hole to high energy levels. This process should be closer to the usual Auger mechanism and differ from EEA in the sense that it implies exciton dissociation into free electron and hole. The promoted carrier could then gain enough energy to be ejected from the nanotube, in its vicinity or at trapping sites on the surface, following a mechanism that has been largely identified for semiconductor nanocrystals~\cite{Masumoto_1996, Valenta_1998}. The other carrier remaining in the nanotube is thus not subject to recombination. It may localize at local fluctuations of the electrostatic potential along the nanotube and be involved in the binding with an exciton to form a trion. 

\section{{Dynamics of the $X^*$ feature and carrier population decay}}
The decay of the $X^*$ IA feature at 1.08~eV, shown in Fig. \ref{FigTrion}b, reflects the dynamics of the population of charge carriers that make the ground state of the single carrier-trion transition. After 900~ps, about 30\% of the maximum transient signal still remains. This slow decay indicates a long-lived feature~\cite{siitonen_2012}, which cannot be fully resolved here since our set-up is limited to pump-probe delays below 1~ns. This timescale is much longer than the lifetime of $S_{11}$-excitons~\cite{hagen_exponential_2005,berciaud_luminescence_2008, gokus_mono-_2010}. This strengthens the hypothesis of carrier localization at trapping sites where they are protected from collisions and their recombination is slowed down~\cite{Santos2011}. Nonetheless, no signal is observed at negative time delays, as seen in Fig.~\ref{FigPeaks}c. This proves that SWCNTs  have relaxed down to their neutral ground state between two consecutive pump pulses. Thus the inverse of the repetition rate of our laser system sets an upper bound of 5~$\mu$s for the lifetime of the trapped charges.

\section{{Conclusion}}
\begin{figure}[!t]
\begin{center}
\includegraphics[width=8.5cm]{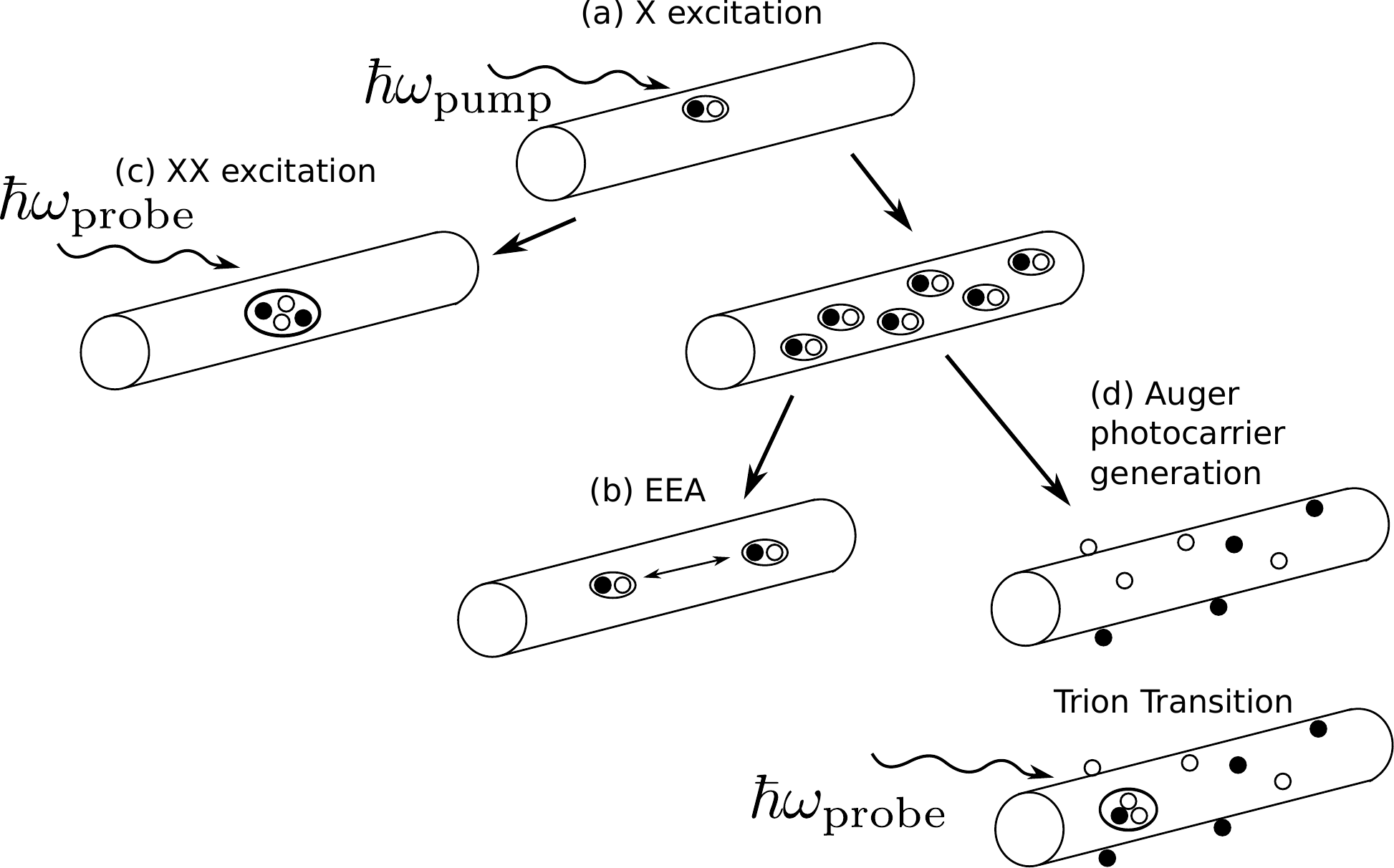}
\caption{Schemes summarizing the different processes involving multi-carrier states in SWCNTS. (a) exciton generation by light absorption, (b) EEA process due to exciton-exciton collisions,(c) biexciton formation from an exciton population by one-photon absorption, (d) Auger carrier photogeneration  and trion excitation. \label{FigMechanism}
}
\end{center}
\end{figure}

Our study uncovers the many-body processes that arise from Coulomb interactions in (6,5) semiconducting SWCNTs, as well as their binding energies and the timescales associated with their dynamics. 
Figure \ref{FigMechanism} illustrates the mechanisms that can lead to the formation of many-body bound states in the case of excitation by a pair of time-delayed, ultrashort pump and probe pulses.
A population of bright singlet $S_{11}$ excitons can be created upon resonant excitation or following ultrafast intersubband decay of higher order excitons (e.g. $S_{22}$ excitons) created with higher photon energies (Fig. \ref{FigMechanism} (a)). Strong Coulomb interactions give rise to exciton-exciton annihilation, which results in a fast decay of the exciton population (Fig. \ref{FigMechanism} (b)). While a significant number of excitons exists, the probe photons may form biexcitons (more generally multi-exciton complexes, cf. Fig. \ref{FigMechanism} (c)). This process results in an induced absorption feature that is red shifted with respect to the bright exciton line by a biexciton binding energy of 130 meV for (6,5) SWCNTs, in good agreement with theoretical predictions. This process obviously competes with EEA. Biexciton formation remains possible as long as there remains a population of excitons. Thus, the temporal dynamics of the biexcitonic feature reflects exciton dynamics. 

For a large exciton density, exciton-exciton interaction predominantly results in Auger recombination, rather than exciton-exciton annihilation, providing enough energy to dissociate an exciton and to expel a single carrier in the vicinity of the nanotube. The carrier remaining in the SWCNT can bind to another e-h pair to form a trion that remain localized on the same sites (Fig. \ref{FigMechanism} (d)). This is evidenced in our study by another induced absorption feature that is red shifted with respect to the bright exciton line by a binding energy of 190 meV for (6,5) SWCNTs. This feature should be observed as long as a population of charge carriers persists in the nanotube. The characteristic timescales for expelled carriers to return into the nanotube are related to the distribution of trapping energies along the nanotube and its dielectric environment. Our measurements indicate that the de-trapping times are equal or larger than the measured lifetime of $\sim 2~ns$ and smaller than $\rm 5\mu s$. With the prospect of designing opto-electronic and photovoltaic nanotube-based devices, our work provides important insights into carrier photogeneration, and possibly carrier multiplication~\cite{GaborScience2009,WangNanolett2010}.

\section*{Acknowledgments}
This work was funded by the "Agence Nationale de la Recherche"

\bibliography{Biblio-Trions}

\begin{thebibliography}{46}
\expandafter\ifx\csname natexlab\endcsname\relax\def\natexlab#1{#1}\fi
\expandafter\ifx\csname bibnamefont\endcsname\relax
  \def\bibnamefont#1{#1}\fi
\expandafter\ifx\csname bibfnamefont\endcsname\relax
  \def\bibfnamefont#1{#1}\fi
\expandafter\ifx\csname citenamefont\endcsname\relax
  \def\citenamefont#1{#1}\fi
\expandafter\ifx\csname url\endcsname\relax
  \def\url#1{\texttt{#1}}\fi
\expandafter\ifx\csname urlprefix\endcsname\relax\def\urlprefix{URL }\fi
\providecommand{\bibinfo}[2]{#2}
\providecommand{\eprint}[2][]{\url{#2}}

\bibitem[{\citenamefont{Wang et~al.}(2005)\citenamefont{Wang, Dukovic, Brus,
  and Heinz}}]{wang_optical_2005}
\bibinfo{author}{\bibfnamefont{F.}~\bibnamefont{Wang}},
  \bibinfo{author}{\bibfnamefont{G.}~\bibnamefont{Dukovic}},
  \bibinfo{author}{\bibfnamefont{L.~E.} \bibnamefont{Brus}}, \bibnamefont{and}
  \bibinfo{author}{\bibfnamefont{T.~F.} \bibnamefont{Heinz}},
  \bibinfo{journal}{Science} \textbf{\bibinfo{volume}{308}},
  \bibinfo{pages}{838} (\bibinfo{year}{2005}),
  \urlprefix\url{http://www.sciencemag.org/cgi/content/abstract/308/5723/838}.

\bibitem[{\citenamefont{Maultzsch et~al.}(2005)\citenamefont{Maultzsch,
  Pomraenke, Reich, Chang, Prezzi, Ruini, Molinari, Strano, Thomsen, and
  Lienau}}]{maultzsch_exciton_2005}
\bibinfo{author}{\bibfnamefont{J.}~\bibnamefont{Maultzsch}},
  \bibinfo{author}{\bibfnamefont{R.}~\bibnamefont{Pomraenke}},
  \bibinfo{author}{\bibfnamefont{S.}~\bibnamefont{Reich}},
  \bibinfo{author}{\bibfnamefont{E.}~\bibnamefont{Chang}},
  \bibinfo{author}{\bibfnamefont{D.}~\bibnamefont{Prezzi}},
  \bibinfo{author}{\bibfnamefont{A.}~\bibnamefont{Ruini}},
  \bibinfo{author}{\bibfnamefont{E.}~\bibnamefont{Molinari}},
  \bibinfo{author}{\bibfnamefont{M.~S.} \bibnamefont{Strano}},
  \bibinfo{author}{\bibfnamefont{C.}~\bibnamefont{Thomsen}}, \bibnamefont{and}
  \bibinfo{author}{\bibfnamefont{C.}~\bibnamefont{Lienau}},
  \bibinfo{journal}{Physical Review B} \textbf{\bibinfo{volume}{72}},
  \bibinfo{pages}{241402} (\bibinfo{year}{2005}),
  \urlprefix\url{http://link.aps.org/doi/10.1103/PhysRevB.72.241402}.

\bibitem[{\citenamefont{Capaz et~al.}(2006)\citenamefont{Capaz, Spataru,
  {Ismail-Beigi}, and Louie}}]{capaz_diameter_2006}
\bibinfo{author}{\bibfnamefont{R.~B.} \bibnamefont{Capaz}},
  \bibinfo{author}{\bibfnamefont{C.~D.} \bibnamefont{Spataru}},
  \bibinfo{author}{\bibfnamefont{S.}~\bibnamefont{{Ismail-Beigi}}},
  \bibnamefont{and} \bibinfo{author}{\bibfnamefont{S.~G.} \bibnamefont{Louie}},
  \bibinfo{journal}{Physical Review B} \textbf{\bibinfo{volume}{74}},
  \bibinfo{pages}{121401} (\bibinfo{year}{2006}),
  \urlprefix\url{http://link.aps.org/doi/10.1103/PhysRevB.74.121401}.

\bibitem[{\citenamefont{Lefebvre and Finnie}(2008)}]{lefebvre_excited_2008}
\bibinfo{author}{\bibfnamefont{J.}~\bibnamefont{Lefebvre}} \bibnamefont{and}
  \bibinfo{author}{\bibfnamefont{P.}~\bibnamefont{Finnie}},
  \bibinfo{journal}{Nano Letters} \textbf{\bibinfo{volume}{8}},
  \bibinfo{pages}{1890} (\bibinfo{year}{2008}),
  \urlprefix\url{http://dx.doi.org/10.1021/nl080518h}.

\bibitem[{\citenamefont{Bachilo et~al.}(2002)\citenamefont{Bachilo, Strano,
  Kittrell, Hauge, Smalley, and Weisman}}]{bachilo_2002}
\bibinfo{author}{\bibfnamefont{S.~M.} \bibnamefont{Bachilo}},
  \bibinfo{author}{\bibfnamefont{M.~S.} \bibnamefont{Strano}},
  \bibinfo{author}{\bibfnamefont{C.}~\bibnamefont{Kittrell}},
  \bibinfo{author}{\bibfnamefont{R.~H.} \bibnamefont{Hauge}},
  \bibinfo{author}{\bibfnamefont{R.~E.} \bibnamefont{Smalley}},
  \bibnamefont{and} \bibinfo{author}{\bibfnamefont{R.~B.}
  \bibnamefont{Weisman}}, \bibinfo{journal}{Science}
  \textbf{\bibinfo{volume}{298}}, \bibinfo{pages}{2361} (\bibinfo{year}{2002}).

\bibitem[{\citenamefont{Dresselhaus et~al.}(2007)\citenamefont{Dresselhaus,
  Dresselhaus, Saito, and Jorio}}]{dresselhaus_exciton_2007}
\bibinfo{author}{\bibfnamefont{M.~S.} \bibnamefont{Dresselhaus}},
  \bibinfo{author}{\bibfnamefont{G.}~\bibnamefont{Dresselhaus}},
  \bibinfo{author}{\bibfnamefont{R.}~\bibnamefont{Saito}}, \bibnamefont{and}
  \bibinfo{author}{\bibfnamefont{A.}~\bibnamefont{Jorio}},
  \bibinfo{journal}{Annual Review of Physical Chemistry}
  \textbf{\bibinfo{volume}{58}}, \bibinfo{pages}{719} (\bibinfo{year}{2007}),
  ISSN \bibinfo{issn}{{0066-426X}},
  \urlprefix\url{http://www.annualreviews.org/doi/abs/10.1146/annurev.physchem.58.032806.104628}.

\bibitem[{\citenamefont{Kammerlander et~al.}(2007)\citenamefont{Kammerlander,
  Prezzi, Goldoni, Molinari, and Hohenester}}]{kammerlander_biexciton_2007}
\bibinfo{author}{\bibfnamefont{D.}~\bibnamefont{Kammerlander}},
  \bibinfo{author}{\bibfnamefont{D.}~\bibnamefont{Prezzi}},
  \bibinfo{author}{\bibfnamefont{G.}~\bibnamefont{Goldoni}},
  \bibinfo{author}{\bibfnamefont{E.}~\bibnamefont{Molinari}}, \bibnamefont{and}
  \bibinfo{author}{\bibfnamefont{U.}~\bibnamefont{Hohenester}},
  \bibinfo{journal}{Physical Review Letters} \textbf{\bibinfo{volume}{99}},
  \bibinfo{pages}{126806} (\bibinfo{year}{2007}),
  \urlprefix\url{http://link.aps.org/doi/10.1103/PhysRevLett.99.126806}.

\bibitem[{\citenamefont{Colombier et~al.}(2012)\citenamefont{Colombier, Selles,
  Rousseau, Lauret, Vialla, Voisin, and Cassabois}}]{colombier_2012}
\bibinfo{author}{\bibfnamefont{L.}~\bibnamefont{Colombier}},
  \bibinfo{author}{\bibfnamefont{J.}~\bibnamefont{Selles}},
  \bibinfo{author}{\bibfnamefont{E.}~\bibnamefont{Rousseau}},
  \bibinfo{author}{\bibfnamefont{J.~S.} \bibnamefont{Lauret}},
  \bibinfo{author}{\bibfnamefont{F.}~\bibnamefont{Vialla}},
  \bibinfo{author}{\bibfnamefont{C.}~\bibnamefont{Voisin}}, \bibnamefont{and}
  \bibinfo{author}{\bibfnamefont{G.}~\bibnamefont{Cassabois}},
  \bibinfo{journal}{Physical Review Letters} \textbf{\bibinfo{volume}{109}},
  \bibinfo{pages}{197402} (\bibinfo{year}{2012}).

\bibitem[{\citenamefont{Pedersen et~al.}(2005)\citenamefont{Pedersen, Pedersen,
  Cornean, and Duclos}}]{pedersen_2005}
\bibinfo{author}{\bibfnamefont{T.~G.} \bibnamefont{Pedersen}},
  \bibinfo{author}{\bibfnamefont{K.}~\bibnamefont{Pedersen}},
  \bibinfo{author}{\bibfnamefont{H.~D.} \bibnamefont{Cornean}},
  \bibnamefont{and} \bibinfo{author}{\bibfnamefont{P.}~\bibnamefont{Duclos}},
  \bibinfo{journal}{Nano Letters} \textbf{\bibinfo{volume}{5}},
  \bibinfo{pages}{291} (\bibinfo{year}{2005}),
  \eprint{http://pubs.acs.org/doi/pdf/10.1021/nl048108q},
  \urlprefix\url{http://pubs.acs.org/doi/abs/10.1021/nl048108q}.

\bibitem[{\citenamefont{Watanabe and Asano}(2011)}]{watanabe_2011}
\bibinfo{author}{\bibfnamefont{K.}~\bibnamefont{Watanabe}} \bibnamefont{and}
  \bibinfo{author}{\bibfnamefont{K.}~\bibnamefont{Asano}},
  \bibinfo{journal}{Phys. Rev. B} \textbf{\bibinfo{volume}{83}},
  \bibinfo{pages}{115406} (\bibinfo{year}{2011}),
  \urlprefix\url{http://link.aps.org/doi/10.1103/PhysRevB.83.115406}.

\bibitem[{\citenamefont{Ronnow et~al.}(2010)\citenamefont{Ronnow, Pedersen, and
  Cornean}}]{ronnow_correlation_2010}
\bibinfo{author}{\bibfnamefont{T.~F.} \bibnamefont{Ronnow}},
  \bibinfo{author}{\bibfnamefont{T.~G.} \bibnamefont{Pedersen}},
  \bibnamefont{and} \bibinfo{author}{\bibfnamefont{H.~D.}
  \bibnamefont{Cornean}}, \bibinfo{journal}{Physical Review B}
  \textbf{\bibinfo{volume}{81}}, \bibinfo{pages}{205446}
  (\bibinfo{year}{2010}),
  \urlprefix\url{http://link.aps.org/doi/10.1103/PhysRevB.81.205446}.

\bibitem[{\citenamefont{Watanabe and Asano}(2012)}]{watanabe_trion_2012}
\bibinfo{author}{\bibfnamefont{K.}~\bibnamefont{Watanabe}} \bibnamefont{and}
  \bibinfo{author}{\bibfnamefont{K.}~\bibnamefont{Asano}},
  \bibinfo{journal}{Phys. Rev. B} \textbf{\bibinfo{volume}{85}},
  \bibinfo{pages}{035416} (\bibinfo{year}{2012}),
  \urlprefix\url{http://link.aps.org/doi/10.1103/PhysRevB.85.035416}.

\bibitem[{\citenamefont{L{\'e}vy et~al.}(1985)\citenamefont{L{\'e}vy,
  H{\"o}nerlage, and Grun}}]{Levy_1985}
\bibinfo{author}{\bibfnamefont{R.}~\bibnamefont{L{\'e}vy}},
  \bibinfo{author}{\bibfnamefont{B.}~\bibnamefont{H{\"o}nerlage}},
  \bibnamefont{and} \bibinfo{author}{\bibfnamefont{J.-B.} \bibnamefont{Grun}},
  \bibinfo{journal}{Physical Review B} \textbf{\bibinfo{volume}{19}},
  \bibinfo{pages}{2326} (\bibinfo{year}{1985}).

\bibitem[{\citenamefont{Bawendi et~al.}(1990)\citenamefont{Bawendi, Wilson,
  Rothberg, Carroll, Jedju, Steigerwald, and Brus}}]{BawendiPRL1990}
\bibinfo{author}{\bibfnamefont{M.~G.} \bibnamefont{Bawendi}},
  \bibinfo{author}{\bibfnamefont{W.~L.} \bibnamefont{Wilson}},
  \bibinfo{author}{\bibfnamefont{L.}~\bibnamefont{Rothberg}},
  \bibinfo{author}{\bibfnamefont{P.~J.} \bibnamefont{Carroll}},
  \bibinfo{author}{\bibfnamefont{T.~M.} \bibnamefont{Jedju}},
  \bibinfo{author}{\bibfnamefont{M.~L.} \bibnamefont{Steigerwald}},
  \bibnamefont{and} \bibinfo{author}{\bibfnamefont{L.~E.} \bibnamefont{Brus}},
  \bibinfo{journal}{Phys. Rev. Lett.} \textbf{\bibinfo{volume}{65}},
  \bibinfo{pages}{1623} (\bibinfo{year}{1990}),
  \urlprefix\url{http://link.aps.org/doi/10.1103/PhysRevLett.65.1623}.

\bibitem[{\citenamefont{Hu et~al.}(1990)\citenamefont{Hu, Koch, Lindberg,
  Peyghambarian, Pollock, and Abraham}}]{HuPRL1990}
\bibinfo{author}{\bibfnamefont{Y.~Z.} \bibnamefont{Hu}},
  \bibinfo{author}{\bibfnamefont{S.~W.} \bibnamefont{Koch}},
  \bibinfo{author}{\bibfnamefont{M.}~\bibnamefont{Lindberg}},
  \bibinfo{author}{\bibfnamefont{N.}~\bibnamefont{Peyghambarian}},
  \bibinfo{author}{\bibfnamefont{E.~L.} \bibnamefont{Pollock}},
  \bibnamefont{and} \bibinfo{author}{\bibfnamefont{F.~F.}
  \bibnamefont{Abraham}}, \bibinfo{journal}{Phys. Rev. Lett.}
  \textbf{\bibinfo{volume}{64}}, \bibinfo{pages}{1805} (\bibinfo{year}{1990}),
  \urlprefix\url{http://link.aps.org/doi/10.1103/PhysRevLett.64.1805}.

\bibitem[{\citenamefont{Wang et~al.}(2004)\citenamefont{Wang, Dukovic, Knoesel,
  Brus, and Heinz}}]{wang_observation_2004}
\bibinfo{author}{\bibfnamefont{F.}~\bibnamefont{Wang}},
  \bibinfo{author}{\bibfnamefont{G.}~\bibnamefont{Dukovic}},
  \bibinfo{author}{\bibfnamefont{E.}~\bibnamefont{Knoesel}},
  \bibinfo{author}{\bibfnamefont{L.~E.} \bibnamefont{Brus}}, \bibnamefont{and}
  \bibinfo{author}{\bibfnamefont{T.~F.} \bibnamefont{Heinz}},
  \bibinfo{journal}{Physical Review B} \textbf{\bibinfo{volume}{70}},
  \bibinfo{pages}{241403} (\bibinfo{year}{2004}),
  \urlprefix\url{http://link.aps.org/doi/10.1103/PhysRevB.70.241403}.

\bibitem[{\citenamefont{Huang and Krauss}(2006)}]{huang_quantized_2006}
\bibinfo{author}{\bibfnamefont{L.}~\bibnamefont{Huang}} \bibnamefont{and}
  \bibinfo{author}{\bibfnamefont{T.~D.} \bibnamefont{Krauss}},
  \bibinfo{journal}{Physical Review Letters} \textbf{\bibinfo{volume}{96}},
  \bibinfo{pages}{057407} (\bibinfo{year}{2006}),
  \urlprefix\url{http://link.aps.org/doi/10.1103/PhysRevLett.96.057407}.

\bibitem[{\citenamefont{Valkunas et~al.}(2006)\citenamefont{Valkunas, Ma, and
  Fleming}}]{valkunas_exciton-exciton_2006}
\bibinfo{author}{\bibfnamefont{L.}~\bibnamefont{Valkunas}},
  \bibinfo{author}{\bibfnamefont{Y.}~\bibnamefont{Ma}}, \bibnamefont{and}
  \bibinfo{author}{\bibfnamefont{G.~R.} \bibnamefont{Fleming}},
  \bibinfo{journal}{Physical Review B} \textbf{\bibinfo{volume}{73}},
  \bibinfo{pages}{115432} (\bibinfo{year}{2006}),
  \urlprefix\url{http://link.aps.org/doi/10.1103/PhysRevB.73.115432}.

\bibitem[{\citenamefont{Xiao et~al.}(2010)\citenamefont{Xiao, Nhan, Wilson, and
  Fraser}}]{xiao_saturation_2010}
\bibinfo{author}{\bibfnamefont{Y.}~\bibnamefont{Xiao}},
  \bibinfo{author}{\bibfnamefont{T.~Q.} \bibnamefont{Nhan}},
  \bibinfo{author}{\bibfnamefont{M.~W.~B.} \bibnamefont{Wilson}},
  \bibnamefont{and} \bibinfo{author}{\bibfnamefont{J.~M.}
  \bibnamefont{Fraser}}, \bibinfo{journal}{Physical Review Letters}
  \textbf{\bibinfo{volume}{104}}, \bibinfo{pages}{017401}
  (\bibinfo{year}{2010}),
  \urlprefix\url{http://link.aps.org/doi/10.1103/PhysRevLett.104.017401}.

\bibitem[{\citenamefont{Kheng et~al.}(1993)\citenamefont{Kheng, Cox,
  d'~Aubign\'e, Bassani, Saminadayar, and Tatarenko}}]{Kheng1993}
\bibinfo{author}{\bibfnamefont{K.}~\bibnamefont{Kheng}},
  \bibinfo{author}{\bibfnamefont{R.~T.} \bibnamefont{Cox}},
  \bibinfo{author}{\bibfnamefont{M.~Y.} \bibnamefont{d'~Aubign\'e}},
  \bibinfo{author}{\bibfnamefont{F.}~\bibnamefont{Bassani}},
  \bibinfo{author}{\bibfnamefont{K.}~\bibnamefont{Saminadayar}},
  \bibnamefont{and}
  \bibinfo{author}{\bibfnamefont{S.}~\bibnamefont{Tatarenko}},
  \bibinfo{journal}{Phys. Rev. Lett.} \textbf{\bibinfo{volume}{71}},
  \bibinfo{pages}{1752} (\bibinfo{year}{1993}),
  \urlprefix\url{http://link.aps.org/doi/10.1103/PhysRevLett.71.1752}.

\bibitem[{\citenamefont{Finkelstein et~al.}(1995)\citenamefont{Finkelstein,
  Shtrikman, and {Bar-Joseph}}}]{Finkelstein_1995}
\bibinfo{author}{\bibfnamefont{G.}~\bibnamefont{Finkelstein}},
  \bibinfo{author}{\bibfnamefont{H.}~\bibnamefont{Shtrikman}},
  \bibnamefont{and}
  \bibinfo{author}{\bibfnamefont{I.}~\bibnamefont{{Bar-Joseph}}},
  \bibinfo{journal}{Physical Review B} \textbf{\bibinfo{volume}{74}},
  \bibinfo{pages}{976} (\bibinfo{year}{1995}).

\bibitem[{\citenamefont{Finkelstein et~al.}(1996)\citenamefont{Finkelstein,
  Shtrikman, and {Bar-Joseph}}}]{finkelstein_shakeup_1996}
\bibinfo{author}{\bibfnamefont{G.}~\bibnamefont{Finkelstein}},
  \bibinfo{author}{\bibfnamefont{H.}~\bibnamefont{Shtrikman}},
  \bibnamefont{and}
  \bibinfo{author}{\bibfnamefont{I.}~\bibnamefont{{Bar-Joseph}}},
  \bibinfo{journal}{Physical Review B} \textbf{\bibinfo{volume}{53}},
  \bibinfo{pages}{12593} (\bibinfo{year}{1996}),
  \urlprefix\url{http://link.aps.org/doi/10.1103/PhysRevB.53.12593}.

\bibitem[{\citenamefont{Brinkmann et~al.}(1999)\citenamefont{Brinkmann, Kudrna,
  Gilliot, Hönerlage, Arnoult, Cibert, and Tatarenko}}]{brinkmann1999}
\bibinfo{author}{\bibfnamefont{D.}~\bibnamefont{Brinkmann}},
  \bibinfo{author}{\bibfnamefont{J.}~\bibnamefont{Kudrna}},
  \bibinfo{author}{\bibfnamefont{P.}~\bibnamefont{Gilliot}},
  \bibinfo{author}{\bibfnamefont{B.}~\bibnamefont{Hönerlage}},
  \bibinfo{author}{\bibfnamefont{A.}~\bibnamefont{Arnoult}},
  \bibinfo{author}{\bibfnamefont{J.}~\bibnamefont{Cibert}}, \bibnamefont{and}
  \bibinfo{author}{\bibfnamefont{S.}~\bibnamefont{Tatarenko}},
  \bibinfo{journal}{Physical Review B} \textbf{\bibinfo{volume}{60}},
  \bibinfo{pages}{4474} (\bibinfo{year}{1999}).

\bibitem[{\citenamefont{Gilliot et~al.}(1999)\citenamefont{Gilliot, Brinkmann,
  Kudrna, Crégut, Lévy, Arnoult, Cibert, and Tatarenko}}]{gilliot1999}
\bibinfo{author}{\bibfnamefont{P.}~\bibnamefont{Gilliot}},
  \bibinfo{author}{\bibfnamefont{D.}~\bibnamefont{Brinkmann}},
  \bibinfo{author}{\bibfnamefont{J.}~\bibnamefont{Kudrna}},
  \bibinfo{author}{\bibfnamefont{O.}~\bibnamefont{Crégut}},
  \bibinfo{author}{\bibfnamefont{R.}~\bibnamefont{Lévy}},
  \bibinfo{author}{\bibfnamefont{A.}~\bibnamefont{Arnoult}},
  \bibinfo{author}{\bibfnamefont{J.}~\bibnamefont{Cibert}}, \bibnamefont{and}
  \bibinfo{author}{\bibfnamefont{S.}~\bibnamefont{Tatarenko}},
  \bibinfo{journal}{Physical Review B} \textbf{\bibinfo{volume}{60}},
  \bibinfo{pages}{5797} (\bibinfo{year}{1999}).

\bibitem[{\citenamefont{Santos et~al.}(2011)\citenamefont{Santos, Yuma,
  Berciaud, Shaver, Gallart, Gilliot, Cognet, and Lounis}}]{Santos2011}
\bibinfo{author}{\bibfnamefont{S.~M.} \bibnamefont{Santos}},
  \bibinfo{author}{\bibfnamefont{B.}~\bibnamefont{Yuma}},
  \bibinfo{author}{\bibfnamefont{S.}~\bibnamefont{Berciaud}},
  \bibinfo{author}{\bibfnamefont{J.}~\bibnamefont{Shaver}},
  \bibinfo{author}{\bibfnamefont{M.}~\bibnamefont{Gallart}},
  \bibinfo{author}{\bibfnamefont{P.}~\bibnamefont{Gilliot}},
  \bibinfo{author}{\bibfnamefont{L.}~\bibnamefont{Cognet}}, \bibnamefont{and}
  \bibinfo{author}{\bibfnamefont{B.}~\bibnamefont{Lounis}},
  \bibinfo{journal}{Phys. Rev. Lett.} \textbf{\bibinfo{volume}{107}},
  \bibinfo{pages}{187401} (\bibinfo{year}{2011}),
  \urlprefix\url{http://link.aps.org/doi/10.1103/PhysRevLett.107.187401}.

\bibitem[{\citenamefont{Matsunaga et~al.}(2011)\citenamefont{Matsunaga,
  Matsuda, and Kanemitsu}}]{matsunaga_observation_2011}
\bibinfo{author}{\bibfnamefont{R.}~\bibnamefont{Matsunaga}},
  \bibinfo{author}{\bibfnamefont{K.}~\bibnamefont{Matsuda}}, \bibnamefont{and}
  \bibinfo{author}{\bibfnamefont{Y.}~\bibnamefont{Kanemitsu}},
  \bibinfo{journal}{Physical Review Letters} \textbf{\bibinfo{volume}{106}},
  \bibinfo{pages}{037404} (\bibinfo{year}{2011}),
  \urlprefix\url{http://link.aps.org/doi/10.1103/PhysRevLett.106.037404}.

\bibitem[{\citenamefont{Ghosh et~al.}(2010)\citenamefont{Ghosh, Bachilo, and
  Weisman}}]{ghosh_advanced_2010}
\bibinfo{author}{\bibfnamefont{S.}~\bibnamefont{Ghosh}},
  \bibinfo{author}{\bibfnamefont{S.~M.} \bibnamefont{Bachilo}},
  \bibnamefont{and} \bibinfo{author}{\bibfnamefont{R.~B.}
  \bibnamefont{Weisman}}, \bibinfo{journal}{Nat Nano}
  \textbf{\bibinfo{volume}{5}}, \bibinfo{pages}{443} (\bibinfo{year}{2010}),
  ISSN \bibinfo{issn}{1748-3387},
  \urlprefix\url{http://dx.doi.org/10.1038/nnano.2010.68}.

\bibitem[{\citenamefont{Manzoni et~al.}(2005)\citenamefont{Manzoni, Gambetta,
  Menna, Meneghetti, Lanzani, and Cerullo}}]{manzoni_intersubband_2005}
\bibinfo{author}{\bibfnamefont{C.}~\bibnamefont{Manzoni}},
  \bibinfo{author}{\bibfnamefont{A.}~\bibnamefont{Gambetta}},
  \bibinfo{author}{\bibfnamefont{E.}~\bibnamefont{Menna}},
  \bibinfo{author}{\bibfnamefont{M.}~\bibnamefont{Meneghetti}},
  \bibinfo{author}{\bibfnamefont{G.}~\bibnamefont{Lanzani}}, \bibnamefont{and}
  \bibinfo{author}{\bibfnamefont{G.}~\bibnamefont{Cerullo}},
  \bibinfo{journal}{Physical Review Letters} \textbf{\bibinfo{volume}{94}},
  \bibinfo{pages}{207401} (\bibinfo{year}{2005}),
  \urlprefix\url{http://link.aps.org/doi/10.1103/PhysRevLett.94.207401}.

\bibitem[{\citenamefont{Shah}(1996, 1999)}]{Shah1999}
\bibinfo{author}{\bibfnamefont{J.}~\bibnamefont{Shah}},
  \emph{\bibinfo{title}{Ultrafast Spectroscopy of Semiconductors and
  Semiconductors Nanostrutures}} (\bibinfo{publisher}{Springer},
  \bibinfo{address}{Berlin}, \bibinfo{year}{1996, 1999}).

\bibitem[{\citenamefont{Haug and Schmitt-Rink}(1984)}]{Haug1984}
\bibinfo{author}{\bibfnamefont{H.}~\bibnamefont{Haug}} \bibnamefont{and}
  \bibinfo{author}{\bibfnamefont{S.}~\bibnamefont{Schmitt-Rink}},
  \bibinfo{journal}{Prog. Quant. Elect.} \textbf{\bibinfo{volume}{9}},
  \bibinfo{pages}{3} (\bibinfo{year}{1984}).

\bibitem[{\citenamefont{Luer et~al.}(2009)\citenamefont{Luer, Hoseinkhani,
  Polli, Crochet, Hertel, and Lanzani}}]{luer_size_2009}
\bibinfo{author}{\bibfnamefont{L.}~\bibnamefont{Luer}},
  \bibinfo{author}{\bibfnamefont{S.}~\bibnamefont{Hoseinkhani}},
  \bibinfo{author}{\bibfnamefont{D.}~\bibnamefont{Polli}},
  \bibinfo{author}{\bibfnamefont{J.}~\bibnamefont{Crochet}},
  \bibinfo{author}{\bibfnamefont{T.}~\bibnamefont{Hertel}}, \bibnamefont{and}
  \bibinfo{author}{\bibfnamefont{G.}~\bibnamefont{Lanzani}},
  \bibinfo{journal}{Nat Phys} \textbf{\bibinfo{volume}{5}}, \bibinfo{pages}{54}
  (\bibinfo{year}{2009}), ISSN \bibinfo{issn}{1745-2473},
  \urlprefix\url{http://dx.doi.org/10.1038/nphys1149}.

\bibitem[{\citenamefont{Nguyen et~al.}(2011)\citenamefont{Nguyen, Voisin,
  Roussignol, Roquelet, Lauret, and Cassabois}}]{NguyenPRL2011}
\bibinfo{author}{\bibfnamefont{D.~T.} \bibnamefont{Nguyen}},
  \bibinfo{author}{\bibfnamefont{C.}~\bibnamefont{Voisin}},
  \bibinfo{author}{\bibfnamefont{P.}~\bibnamefont{Roussignol}},
  \bibinfo{author}{\bibfnamefont{C.}~\bibnamefont{Roquelet}},
  \bibinfo{author}{\bibfnamefont{J.~S.} \bibnamefont{Lauret}},
  \bibnamefont{and}
  \bibinfo{author}{\bibfnamefont{G.}~\bibnamefont{Cassabois}},
  \bibinfo{journal}{Phys. Rev. Lett.} \textbf{\bibinfo{volume}{107}},
  \bibinfo{pages}{127401} (\bibinfo{year}{2011}),
  \urlprefix\url{http://link.aps.org/doi/10.1103/PhysRevLett.107.127401}.

\bibitem[{\citenamefont{Zhu et~al.}(2007)\citenamefont{Zhu, Crochet, Arnold,
  Hersam, Ulbricht, Resasco, and Hertel}}]{zhu_pump-probe_2007}
\bibinfo{author}{\bibfnamefont{Z.}~\bibnamefont{Zhu}},
  \bibinfo{author}{\bibfnamefont{J.}~\bibnamefont{Crochet}},
  \bibinfo{author}{\bibfnamefont{M.~S.} \bibnamefont{Arnold}},
  \bibinfo{author}{\bibfnamefont{M.~C.} \bibnamefont{Hersam}},
  \bibinfo{author}{\bibfnamefont{H.}~\bibnamefont{Ulbricht}},
  \bibinfo{author}{\bibfnamefont{D.}~\bibnamefont{Resasco}}, \bibnamefont{and}
  \bibinfo{author}{\bibfnamefont{T.}~\bibnamefont{Hertel}},
  \bibinfo{journal}{The Journal of Physical Chemistry C}
  \textbf{\bibinfo{volume}{111}}, \bibinfo{pages}{3831} (\bibinfo{year}{2007}),
  \urlprefix\url{http://dx.doi.org/10.1021/jp0669411}.

\bibitem[{\citenamefont{Nishihara et~al.}(2012)\citenamefont{Nishihara, Yamada,
  and Katemitsu}}]{Nishihara_2012}
\bibinfo{author}{\bibfnamefont{T.}~\bibnamefont{Nishihara}},
  \bibinfo{author}{\bibfnamefont{Y.}~\bibnamefont{Yamada}}, \bibnamefont{and}
  \bibinfo{author}{\bibfnamefont{Y.}~\bibnamefont{Katemitsu}},
  \bibinfo{journal}{Physical Review B} \textbf{\bibinfo{volume}{86}},
  \bibinfo{pages}{075449} (\bibinfo{year}{2012}).

\bibitem[{\citenamefont{Matsunaga et~al.}(2010)\citenamefont{Matsunaga,
  Matsuda, and Kanemitsu}}]{matsunaga_origin_2010}
\bibinfo{author}{\bibfnamefont{R.}~\bibnamefont{Matsunaga}},
  \bibinfo{author}{\bibfnamefont{K.}~\bibnamefont{Matsuda}}, \bibnamefont{and}
  \bibinfo{author}{\bibfnamefont{Y.}~\bibnamefont{Kanemitsu}},
  \bibinfo{journal}{Physical Review B} \textbf{\bibinfo{volume}{81}},
  \bibinfo{pages}{033401} (\bibinfo{year}{2010}),
  \urlprefix\url{http://link.aps.org/doi/10.1103/PhysRevB.81.033401}.

\bibitem[{\citenamefont{Vora et~al.}(2010)\citenamefont{Vora, Tu, Mele, Zheng,
  and Kikkawa}}]{vora_chirality_2010}
\bibinfo{author}{\bibfnamefont{P.~M.} \bibnamefont{Vora}},
  \bibinfo{author}{\bibfnamefont{X.}~\bibnamefont{Tu}},
  \bibinfo{author}{\bibfnamefont{E.~J.} \bibnamefont{Mele}},
  \bibinfo{author}{\bibfnamefont{M.}~\bibnamefont{Zheng}}, \bibnamefont{and}
  \bibinfo{author}{\bibfnamefont{J.~M.} \bibnamefont{Kikkawa}},
  \bibinfo{journal}{Physical Review B} \textbf{\bibinfo{volume}{81}},
  \bibinfo{pages}{155123} (\bibinfo{year}{2010}),
  \urlprefix\url{http://link.aps.org/doi/10.1103/PhysRevB.81.155123}.

\bibitem[{\citenamefont{Matsuda et~al.}(2008)\citenamefont{Matsuda, Inoue,
  Murakami, Maruyama, and Kanemitsu}}]{matsuda_exciton_2008}
\bibinfo{author}{\bibfnamefont{K.}~\bibnamefont{Matsuda}},
  \bibinfo{author}{\bibfnamefont{T.}~\bibnamefont{Inoue}},
  \bibinfo{author}{\bibfnamefont{Y.}~\bibnamefont{Murakami}},
  \bibinfo{author}{\bibfnamefont{S.}~\bibnamefont{Maruyama}}, \bibnamefont{and}
  \bibinfo{author}{\bibfnamefont{Y.}~\bibnamefont{Kanemitsu}},
  \bibinfo{journal}{Physical Review B} \textbf{\bibinfo{volume}{77}},
  \bibinfo{pages}{033406} (\bibinfo{year}{2008}),
  \urlprefix\url{http://link.aps.org/doi/10.1103/PhysRevB.77.033406}.

\bibitem[{\citenamefont{Park et~al.}(2012)\citenamefont{Park, Hirana, Mouri,
  Miyauchi, Nakashima, and Matsuda}}]{Park_2012}
\bibinfo{author}{\bibfnamefont{J.~S.} \bibnamefont{Park}},
  \bibinfo{author}{\bibfnamefont{Y.}~\bibnamefont{Hirana}},
  \bibinfo{author}{\bibfnamefont{S.}~\bibnamefont{Mouri}},
  \bibinfo{author}{\bibfnamefont{Y.}~\bibnamefont{Miyauchi}},
  \bibinfo{author}{\bibfnamefont{N.}~\bibnamefont{Nakashima}},
  \bibnamefont{and} \bibinfo{author}{\bibfnamefont{K.}~\bibnamefont{Matsuda}},
  \bibinfo{journal}{Journal of the American Chemical Society}
  \textbf{\bibinfo{volume}{0}}, \bibinfo{pages}{14461} (\bibinfo{year}{2012}),
  \eprint{http://pubs.acs.org/doi/pdf/10.1021/ja304282j},
  \urlprefix\url{http://pubs.acs.org/doi/abs/10.1021/ja304282j}.

\bibitem[{\citenamefont{Hagen et~al.}(2005)\citenamefont{Hagen, Steiner,
  Raschke, Lienau, Hertel, Qian, Meixner, and
  Hartschuh}}]{hagen_exponential_2005}
\bibinfo{author}{\bibfnamefont{A.}~\bibnamefont{Hagen}},
  \bibinfo{author}{\bibfnamefont{M.}~\bibnamefont{Steiner}},
  \bibinfo{author}{\bibfnamefont{M.~B.} \bibnamefont{Raschke}},
  \bibinfo{author}{\bibfnamefont{C.}~\bibnamefont{Lienau}},
  \bibinfo{author}{\bibfnamefont{T.}~\bibnamefont{Hertel}},
  \bibinfo{author}{\bibfnamefont{H.}~\bibnamefont{Qian}},
  \bibinfo{author}{\bibfnamefont{A.~J.} \bibnamefont{Meixner}},
  \bibnamefont{and}
  \bibinfo{author}{\bibfnamefont{A.}~\bibnamefont{Hartschuh}},
  \bibinfo{journal}{Physical Review Letters} \textbf{\bibinfo{volume}{95}},
  \bibinfo{pages}{197401} (\bibinfo{year}{2005}),
  \urlprefix\url{http://link.aps.org/doi/10.1103/PhysRevLett.95.197401}.

\bibitem[{\citenamefont{Berciaud et~al.}(2008)\citenamefont{Berciaud, Cognet,
  and Lounis}}]{berciaud_luminescence_2008}
\bibinfo{author}{\bibfnamefont{S.}~\bibnamefont{Berciaud}},
  \bibinfo{author}{\bibfnamefont{L.}~\bibnamefont{Cognet}}, \bibnamefont{and}
  \bibinfo{author}{\bibfnamefont{B.}~\bibnamefont{Lounis}},
  \bibinfo{journal}{Physical Review Letters} \textbf{\bibinfo{volume}{101}},
  \bibinfo{pages}{077402} (\bibinfo{year}{2008}),
  \urlprefix\url{http://link.aps.org/doi/10.1103/PhysRevLett.101.077402}.

\bibitem[{\citenamefont{Gokus et~al.}(2010)\citenamefont{Gokus, Cognet, Duque,
  Pasquali, Hartschuh, and Lounis}}]{gokus_mono-_2010}
\bibinfo{author}{\bibfnamefont{T.}~\bibnamefont{Gokus}},
  \bibinfo{author}{\bibfnamefont{L.}~\bibnamefont{Cognet}},
  \bibinfo{author}{\bibfnamefont{J.~G.} \bibnamefont{Duque}},
  \bibinfo{author}{\bibfnamefont{M.}~\bibnamefont{Pasquali}},
  \bibinfo{author}{\bibfnamefont{A.}~\bibnamefont{Hartschuh}},
  \bibnamefont{and} \bibinfo{author}{\bibfnamefont{B.}~\bibnamefont{Lounis}},
  \bibinfo{journal}{The Journal of Physical Chemistry C}
  \textbf{\bibinfo{volume}{114}}, \bibinfo{pages}{14025}
  (\bibinfo{year}{2010}), \urlprefix\url{http://dx.doi.org/10.1021/jp1049217}.

\bibitem[{\citenamefont{Masumoto}(1996)}]{Masumoto_1996}
\bibinfo{author}{\bibfnamefont{Y.}~\bibnamefont{Masumoto}},
  \bibinfo{journal}{Journal of Luminescence} \textbf{\bibinfo{volume}{70}},
  \bibinfo{pages}{386} (\bibinfo{year}{1996}).

\bibitem[{\citenamefont{Valenta et~al.}(1998)\citenamefont{Valenta, Moniatte,
  Gilliot, Hönerlage, Grun, and Levy}}]{Valenta_1998}
\bibinfo{author}{\bibfnamefont{J.}~\bibnamefont{Valenta}},
  \bibinfo{author}{\bibfnamefont{J.}~\bibnamefont{Moniatte}},
  \bibinfo{author}{\bibfnamefont{P.}~\bibnamefont{Gilliot}},
  \bibinfo{author}{\bibfnamefont{B.}~\bibnamefont{Hönerlage}},
  \bibinfo{author}{\bibfnamefont{J.-B.} \bibnamefont{Grun}}, \bibnamefont{and}
  \bibinfo{author}{\bibfnamefont{R.}~\bibnamefont{Levy}},
  \bibinfo{journal}{Physical Review B} \textbf{\bibinfo{volume}{57}},
  \bibinfo{pages}{1774} (\bibinfo{year}{1998}).

\bibitem[{\citenamefont{Siitonen et~al.}(2012)\citenamefont{Siitonen, Bachilo,
  Tsyboulski, and Weisman}}]{siitonen_2012}
\bibinfo{author}{\bibfnamefont{A.~J.} \bibnamefont{Siitonen}},
  \bibinfo{author}{\bibfnamefont{S.~M.} \bibnamefont{Bachilo}},
  \bibinfo{author}{\bibfnamefont{D.~A.} \bibnamefont{Tsyboulski}},
  \bibnamefont{and} \bibinfo{author}{\bibfnamefont{R.~B.}
  \bibnamefont{Weisman}}, \bibinfo{journal}{Nano Letters}
  \textbf{\bibinfo{volume}{12}}, \bibinfo{pages}{33} (\bibinfo{year}{2012}).

\bibitem[{\citenamefont{Gabor et~al.}(2009)\citenamefont{Gabor, Zhong, Bosnick,
  Park, and McEuen}}]{GaborScience2009}
\bibinfo{author}{\bibfnamefont{N.~M.} \bibnamefont{Gabor}},
  \bibinfo{author}{\bibfnamefont{Z.}~\bibnamefont{Zhong}},
  \bibinfo{author}{\bibfnamefont{K.}~\bibnamefont{Bosnick}},
  \bibinfo{author}{\bibfnamefont{J.}~\bibnamefont{Park}}, \bibnamefont{and}
  \bibinfo{author}{\bibfnamefont{P.~L.} \bibnamefont{McEuen}},
  \bibinfo{journal}{Science} \textbf{\bibinfo{volume}{325}},
  \bibinfo{pages}{1367} (\bibinfo{year}{2009}),
  \eprint{http://www.sciencemag.org/content/325/5946/1367.full.pdf},
  \urlprefix\url{http://www.sciencemag.org/content/325/5946/1367.abstract}.

\bibitem[{\citenamefont{Wang et~al.}(2010)\citenamefont{Wang, Khafizov, Tu,
  Zheng, and Krauss}}]{WangNanolett2010}
\bibinfo{author}{\bibfnamefont{S.}~\bibnamefont{Wang}},
  \bibinfo{author}{\bibfnamefont{M.}~\bibnamefont{Khafizov}},
  \bibinfo{author}{\bibfnamefont{X.}~\bibnamefont{Tu}},
  \bibinfo{author}{\bibfnamefont{M.}~\bibnamefont{Zheng}}, \bibnamefont{and}
  \bibinfo{author}{\bibfnamefont{T.~D.} \bibnamefont{Krauss}},
  \bibinfo{journal}{Nano Letters} \textbf{\bibinfo{volume}{10}},
  \bibinfo{pages}{2381} (\bibinfo{year}{2010}),
  \eprint{http://pubs.acs.org/doi/pdf/10.1021/nl100343j},
  \urlprefix\url{http://pubs.acs.org/doi/abs/10.1021/nl100343j}.

\end{thebibliography}

\end{document}